\documentclass[a4paper,doc]{apa6}

\usepackage[english]{babel}
\usepackage[utf8x]{inputenc}
\usepackage{amsmath}
\usepackage{amssymb}
\usepackage{graphicx}
\usepackage[colorinlistoftodos]{todonotes}
\usepackage{algpseudocode}
\usepackage{natbib}

\usepackage{natbib}
\usepackage{url}

\title{Brief Report on Estimating Regularized Gaussian Networks from Continuous and Ordinal Data}
\shorttitle{Report on Regularized Gaussian Networks}
\author{Sacha Epskamp}
\affiliation{University of Amsterdam: Department of Psychological Methods}

\abstract{
In recent literature, the Gaussian Graphical model \citep[GGM;][]{lauritzen1996graphical}, a network of partial correlation coefficients, has been used to capture potential dynamic relationships between observed variables. The GGM can be estimated using regularization in combination with model selection using the extended Bayesian Information Criterion \citep{foygel2010extended}. I term this methodology \emph{GeLasso}, and asses its performance using a plausible psychological network structure with both continuous and ordinal datasets. Simulation results indicate that GeLasso works well as an out-of-the-box method to estimate network structures.}

\begin{document}
\maketitle

Recent years have seen a emergance of the network conceptualisation of psychology, pyschiatry and health sciences, in which relationships between attitudes, moods, clinical symptoms and other observed variables are seen as interacting components in a dynamical system, rather than indicators of one or more latent constructs \citep{cramer2010comorbidity,borsboom2011small,schmittmann2013deconstructing}. The models proposed take the form of \emph{networks}, in which nodes represent observed variables which are connected by edges representing statistical relationships between these variables \citep{epskamp2012qgraph}. These models strikingly differ from typically used network models such as social networks \citep{wasserman1994social} or transportation networks \citep{newman2010}, in that variables are not static entities (e.g., people or cities) but random variables, and links are not observed (e.g., friendships or roads) but need to be estimated \citep{stability}. 

When data is assumed multivariate normal distributed, a prominent, interpretable and easy to use network model is the \emph{Gaussian Graphical Model} \citep[GGM;][]{lauritzen1996graphical,dynamics}, a network in which edges represent \emph{partial correlations} between two variables after conditioning on all other variables in the network. Such networks are being extensively being applied to psychological datasets \citep[e.g.,][]{mcnally2015mental,kossakowski2015,isvoranu,fried2016,van2015association}. To control for spurious relationships a regularization technique called the `least absolute shrinkage and selection operator' \citep[LASSO;][]{tibshirani1996regression} is often used \citep{costantini2015state}. The graphical LASSO \citep[glasso;][]{friedman2008sparse} is a particularly fast variant of the LASSO that only requires a covariance matrix. As especially psychological data are often ordinal, an estimate of the covariance matrix can be obtained by computing polychoric and polyserial correlations \citep{olsson1979maximum,olsson1982polyserial}, which can be used in the glasso algorithm \citep{tutorial}. For a detailed methodological introduction to the GGM I refer the reader to \citet{dynamics}.

LASSO regularization utilizes a tuning parameter, $\lambda$, which controls the sparsity of the network. Typically, a range of networks is estimated under different values of $\lambda$ (Zhao \& Yu, \citeyear{zhao2006model}). The value for $\lambda$ under which no edges are retained (the empty network), $\lambda_{\mathrm{max}}$, is set to the largest absolute correlation (Zhao et al., \citeyear{huge}). Next, a minimum value can be chosen by multiplying some ratio $R$ (typically set to $0.01$ or $0.1$) with this maximum value:
\[
\lambda_{\mathrm{min}} = R \lambda_{\mathrm{max}}.
\]
A logorithmically spaced range of tuning parameters (typically $100$ different values), ranging from $\lambda_{\mathrm{min}}$ to $\lambda_{\mathrm{max}}$, can be used to estimate different networks. Subsequently, an optimal network with many true connections and few spurious connections can be obtained through model selection \citep{drton2004model}. The network that has the least cross-validation prediction error or the lowest value of some information criterion is often the selected network. The extended Bayesian Information Criterion (EBIC; \citealt{chen2008extended,foygel2010extended}) adds an extra penalty for model complexity to the typical BIC and has been shown to work well in high-dimensional network model selection \citep{foygel2010extended,barber2015high,van2014new}. The EBIC uses a hyperparameter, $\gamma$, wich controls the extra penalization; $\gamma=0$ leads to the EBIC reducing to the BIC, and higher values of $\gamma$ lead to more penalization. Typically, $\gamma$ is set between $0$ and $1$. I will shorten EBIC selection of GGM models using LASSO regularization via the glasso algorithm to \emph{GeLasso}\footnote{The term GeLasso is line with \citet{van2014new}, who use \emph{eLasso} in the context of estimating a pairwise markov random field for binary variables.}.

While GeLasso has already been shown to work well in retrieving the GGM structure \citep[who suggest $\gamma = 0.5$]{foygel2010extended}, it has not been validated in plausible scenarios for psychological networks. In addition, no simulation study has assessed the performance of using a polychoric correlation matrix in this methodology. To this end, this report presents a simulation study that assesses the performance of GeLasso in a plausible psychological network structure. Furthermore, the simulation study varied $R$ and $\gamma$ in order to provide recommendations of these parameters in estimating psychological networks. The simulation study makes use of the \verb|qgraph| package \citep{epskamp2012qgraph}, which implements GeLasso using the \verb|glasso| package for the glasso algorithm \citep{glasso}.

\section{Methods}

\begin{figure}
\centering
\includegraphics[width=1\linewidth]{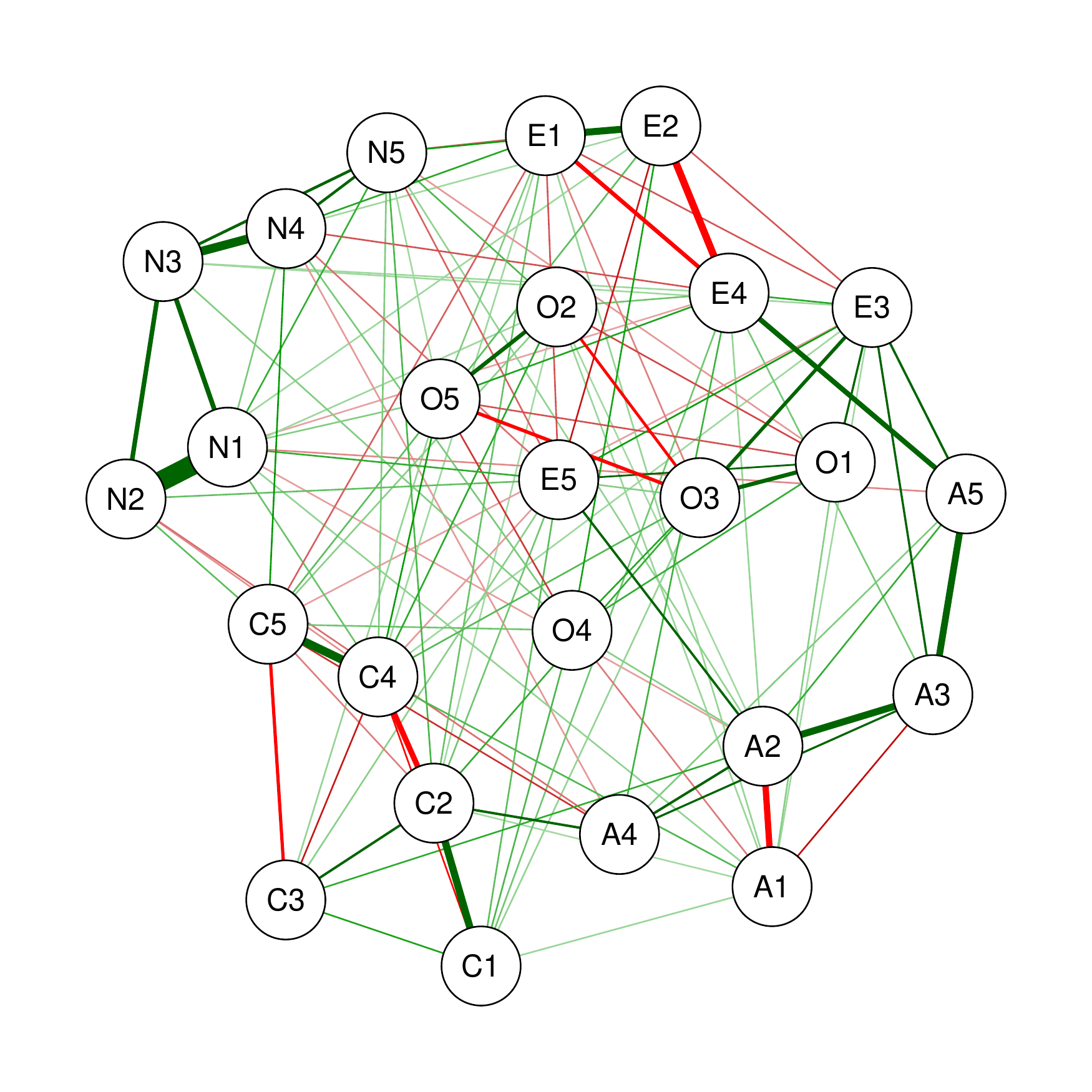}
\caption{True Gaussian graphical model used in simulation study. Nodes represent personality inventory items and edges can be interpreted as partial correlation coefficients. Green edges indicate positive partial correlations, red edges indicate negative partial correlations and the wider and more saturated the edge the stronger the correlation \citep{epskamp2012qgraph}. The network was obtained by computing the (unregularized) sample partial correlation network and removing all absolute edges below 0.05.}
\label{fig:truenetwork}
\end{figure}

To obtain a representative psychological network structure, the \verb|bfi| dataset from the \verb|psych| package \citep{psych} was used on the Big 5 personality traits \citep{benet1998cinco,digman1989five,goldberg2001alternative,goldberg1998structure, mccrae1997personality}. The \verb|bfi| dataset consists of $2{,}800$ observations of $25$ personality inventory items. The network structure was obtained by computing the sample partial correlation coefficients (negative standardized inverse of the sample variance-covariance matrix; \citealt{lauritzen1996graphical}). Next, to create a sparse network all absolute edge weights below $0.05$ were set to zero, thus removing edges from the network. Figure~\ref{fig:truenetwork} shows the resulting network structure. In this network, 125 out of 300 possible edges were nonzero ($41.6\%$). While this network is not the most appropriate network based on this dataset, it functions well as a proxy for psychological network structures as it is both sparse (has missing edges) and has parameter values that are not shrunken by the LASSO.

In the simulation study, data was generated based on the network of Figure~\ref{fig:truenetwork}. Following, the network was estimated using the \verb|EBICglasso| function in the \verb|qgraph| package \citep{epskamp2012qgraph}. Sample size was varied between $50$, $100$, $250$, $500$, $1{,}000$, and $2{,}500$, $\gamma$ was varied between $0$, $0.25$, $0.5$, $0.75$, and $1$, and $R$ was varied between $0.001$, $0.01$ and $0.1$. The data was either simulated to be multivariate normal, in which case Pearson correlations were used in estimation, or ordinal, in which case polychoric correlations were used in the estimation. Ordinal data was created by sampling four thresholds for every variable from the standard normal distribution, and next using these thresholds to cut each variable in five levels. To compute polychoric correlations, the \verb|cor_auto| function was used, which uses the \verb|lavCor| function of the \verb|lavaan| package \citep{lavaan}. The number of different $\lambda$ values used in generating networks was set to $100$ (the default in \verb|qgraph|).

For each simulation, in addition to the correlation between estimated and true edge weights, the sensetivity and specificity were computed \citep{van2014new,tutorial}. The \emph{sensitivity}, also termed the true-positive rate, indicates the proportion of edges in the true network that were estimated to be nonzero:
\[
\text{sensitivity} = \frac{\text{\# true positives}}{\text{\# true positives} + \text{\# of false negatives}}. 
\]
Specificity, also termed the true negative rate, indicates the proportion of true missing edges that were also estimated to be missing:
\[
\text{specificity} = \frac{\text{\# true negatives}}{\text{\# true negatives} + \text{\# false positives}}.
\]
When specificity is high, there are not many false positives (edges detected to be nonzero that are zero in the true network) in the estimated network.

\section{Results}

\begin{figure*}
\centering
\includegraphics[width=1\linewidth]{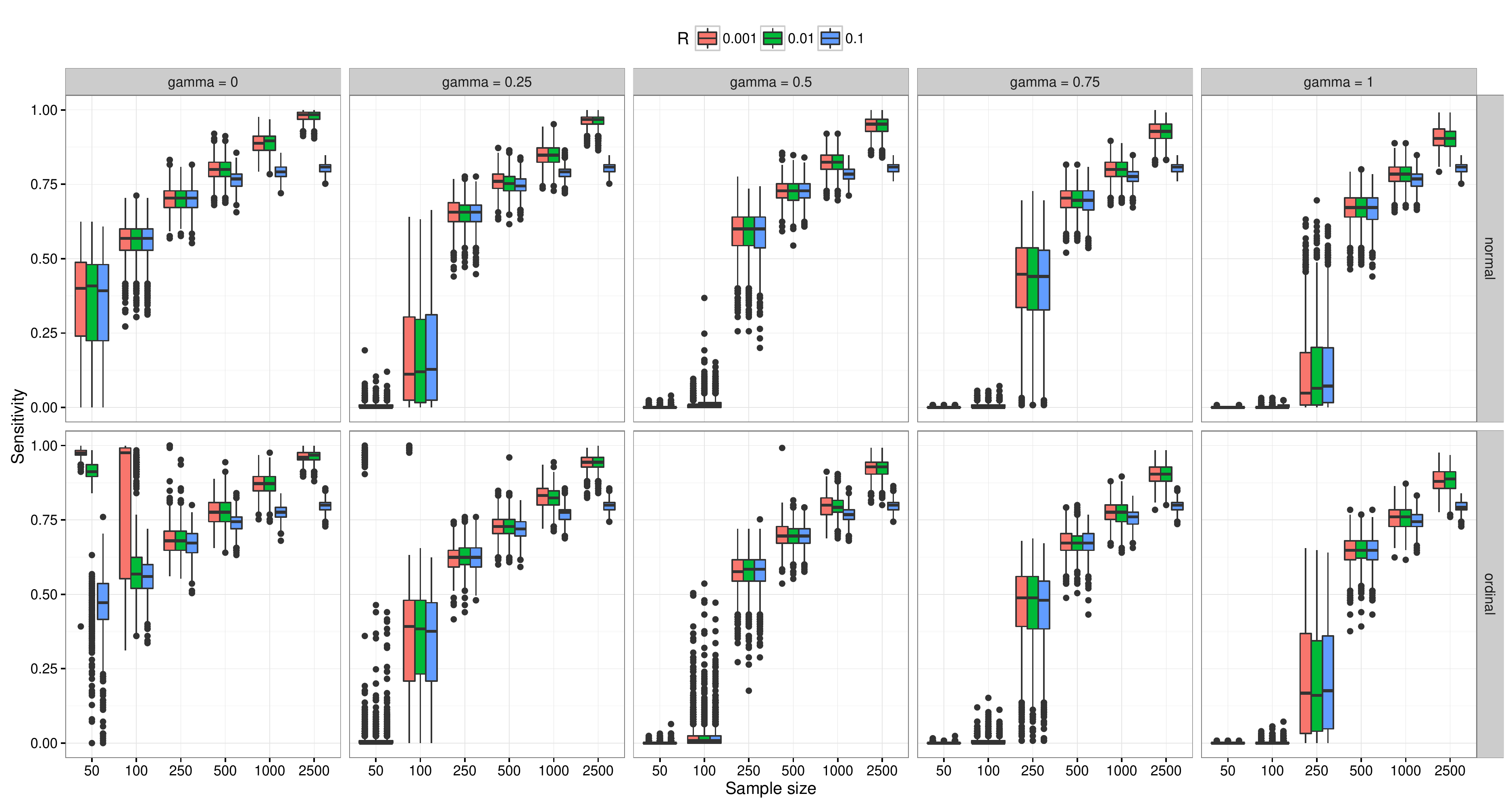}
\caption{Sensitivity of the simulated datasets. Data is represented in standard boxplots \citep{mcgill1978variations}. Horizontal panels indicate different EBIC hyperparameter values, vertical panels indicate if data was normal (Pearson correlations) or ordinal (polychoric correlations) and the color of the boxplots indicate the different ratio values used in setting the LASSO tuning parameter range. When sensitivity is high, true edges are likely to be detected.}
\label{fig:sensitivity}
\end{figure*}

Each of the conditions was replicated $1{,}000$ times, leading to $180{,}000$ simulated datasets. Figure~\ref{fig:sensitivity} shows the sensitivity of the analyses. This figure shows that sensitivity increases with sample size and is high for large sample sizes. When $\gamma > 0$, small sample sizes are likely to result in empty networks (no edges), indicating a sensitivity of 0. When ordinal data is used, small sample sizes ($50$ and $100$) resulted in far too densely connected networks that are hard to interpret. Setting $\gamma$ to be higher remediated this by estimating empty networks. At higher sample sizes, $\gamma$ does not play a role and sensitivity is comparable in all conditions. Using $R = 0.1$ remediates the poor performance of polychoric correlations in lower sample sizes, but also creates an upper bound to sensitivity at higher sample sizes.

\begin{figure*}
\centering
\includegraphics[width=1\linewidth]{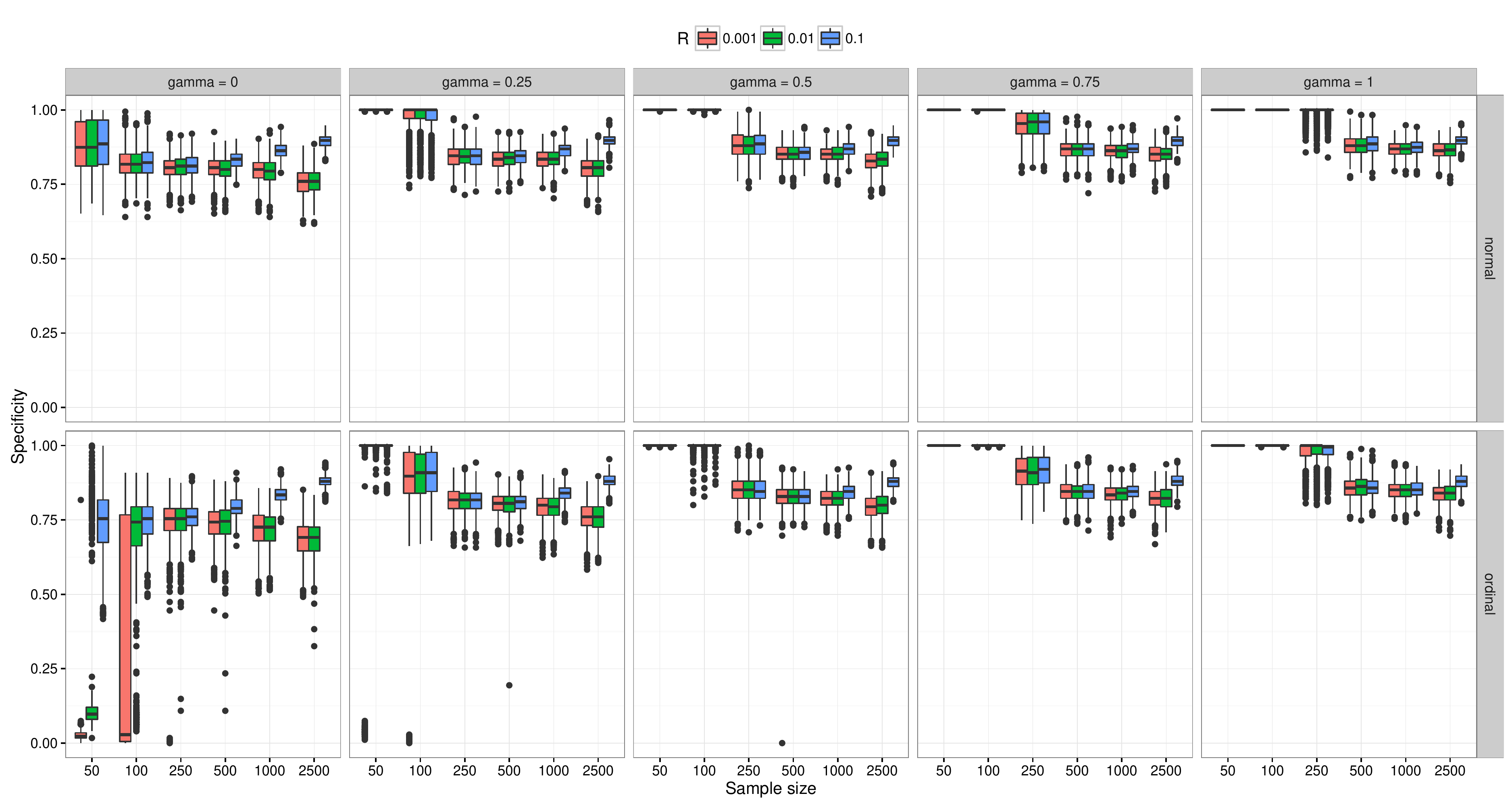}
\caption{The specificity of the simulated datasets. When specificity is high, there are not many edges in the estimated network that are not present in the true network. See caption of Figure~\ref{fig:sensitivity} for more details.}
\label{fig:specificity}
\end{figure*}

\begin{figure*}
\centering
\includegraphics[width=1\linewidth]{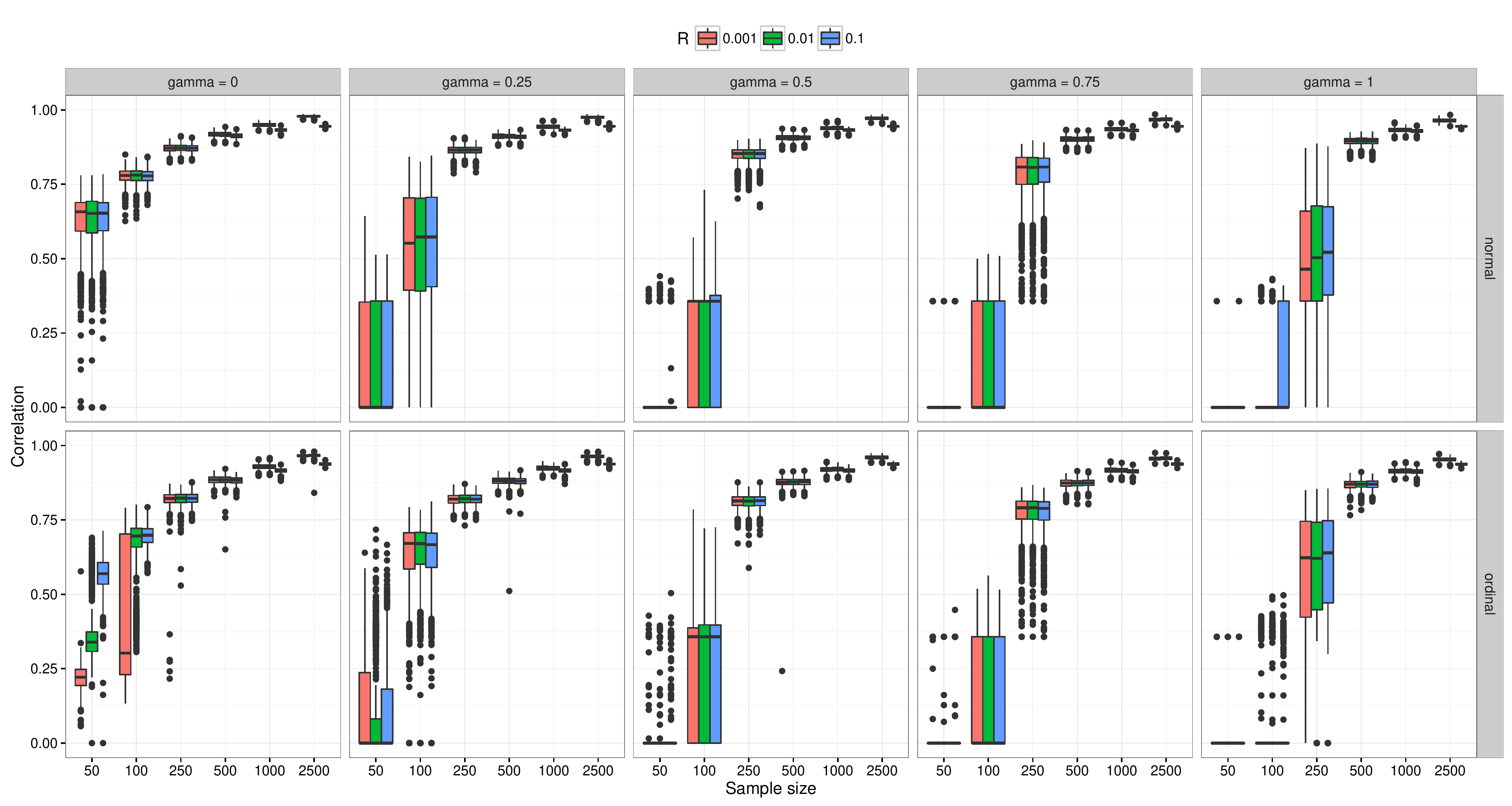}
\caption{Correlation between true edge weights and estimated edge weights.  See caption of Figure~\ref{fig:sensitivity} for more details.}
\label{fig:correlation}
\end{figure*}

Figure~\ref{fig:specificity} shows the specificity of the analyses, which was all-around high except for the lower sample sizes in ordinal data using $R = 0.01$ or $R = 0.001$. Some outliers indicate that fully connected networks were estimated in ordinal data even when setting $\gamma = 0.25$ in small sample sizes. In all other conditions specificity was comparably high, with higher $\gamma$ values only performing slightly better. Figure~\ref{fig:correlation} shows the correlation between true and estimated edge weights. This figure shows a comparable good performance from sample sizes of $250$ and higher in all conditions, with $\gamma$ values up to $0.5$ outperforming the higher $\gamma$ values. It should be noted that the correlation was set to zero if the estimated network had no edges (all edge weights were then zero). 

\section{Conclusion}

In this brief report I assessed the performance of GeLasso in $180{,}000$ simulated datasets using a plausible psychological network structure. Results indicate that GeLasso performs well in estimating psychological networks using both Pearson correlations or polychoric correlations. The default setup of \verb|qgraph| uses $\gamma=0.5$ and $R = 0.01$, which are shown to work well in all conditions. Setting $\gamma = 0.25$ improved the detection rate, but sometimes led to poorly estimated networks based on polychoric correlations. $\gamma$ can be set to $0$ to err more on the side of discovery \citep{dziak2012sensitivity}, but should be done with care in low sample polychoric correlation matrices. All conditions showed increasing sensitivity with sample size and a high specificity all-around. This is comparable to other network estimation techniques \citep{van2014new}, and shows that even though a network does not contain all true edges, the edges that are returned can usually be expected to be genuine. The high correlation furthermore indicated that the strongest true edges are usually estimated to be strong as well. 

The estimation of psychological networks is a rapidly evolving field of research. In addition to the GeLasso method many other network analysis methods exists (e.g., Zhao et al., \citeyear{huge}; \citealt{parcor}; \citealt{kalisch2012causal}). When variables are binary, a more appropriate model to use is the Ising Model \citep{van2014new}. In addition, new and promising methods have been developed for estimating network structures with mixed continous and catagorical variables \citep{mgm}. For a  tutorial on both using the GeLasso method and on assessing the stability of such network structures, I refer the reader to \citet{tutorial}. 

\section{Acknowledgements}

I would like to thank Lourens J.\ Waldorp, Eiko I.\ Fried and Adela M.\ Isvoranu for their helpful comments and feedback.

\bibliographystyle{apalike}
\bibliography{Bibliography}

\begin{thebibliography}{}

\bibitem[Barber et~al., 2015]{barber2015high}
Barber, R.~F., Drton, M., and Others (2015).
\newblock {High-dimensional Ising model selection with Bayesian information
  criteria}.
\newblock {\em Electronic Journal of Statistics}, 9(1):567--607.

\bibitem[Benet-Martinez and John, 1998]{benet1998cinco}
Benet-Martinez, V. and John, O. (1998).
\newblock {Los Cinco Grandes across Cultures and Ethnic Groups: Multitrait
  Multimethod Analyses of the Big Five in Spanish and English}.
\newblock {\em Journal of Personality and Social Psychology}, 75:729--750.

\bibitem[Borsboom et~al., 2011]{borsboom2011small}
Borsboom, D., Cramer, A. O.~J., Schmittmann, V.~D., Epskamp, S., and Waldorp,
  L.~J. (2011).
\newblock The small world of psychopathology.
\newblock {\em PloS one}, 6(11):e27407.

\bibitem[Costantini et~al., 2015]{costantini2015state}
Costantini, G., Epskamp, S., Borsboom, D., Perugini, M., M\~{o}ttus, R.,
  Waldorp, L.~J., and Cramer, A. O.~J. (2015).
\newblock State of the {aRt} personality research: A tutorial on network
  analysis of personality data in {R}.
\newblock {\em Journal of Research in Personality}, 54:13--29.

\bibitem[Cramer et~al., 2010]{cramer2010comorbidity}
Cramer, A. O.~J., Waldorp, L., van~der Maas, H., and Borsboom, D. (2010).
\newblock {Comorbidity: A Network Perspective}.
\newblock {\em Behavioral and Brain Sciences}, 33(2-3):137--150.

\bibitem[Digman, 1989]{digman1989five}
Digman, J. (1989).
\newblock {Five Robust Trait Dimensions: Development, Stability, and Utility}.
\newblock {\em Journal of Personality}, 57(2):195--214.

\bibitem[Drton and Perlman, 2004]{drton2004model}
Drton, M. and Perlman, M.~D. (2004).
\newblock Model selection for gaussian concentration graphs.
\newblock {\em Biometrika}, 91(3):591--602.

\bibitem[Dziak et~al., 2012]{dziak2012sensitivity}
Dziak, J.~J., Coffman, D.~L., Lanza, S.~T., and Li, R. (2012).
\newblock Sensitivity and specificity of information criteria.
\newblock {\em The Methodology Center and Department of Statistics, Penn State,
  The Pennsylvania State University}.

\bibitem[Epskamp et~al., 2016]{stability}
Epskamp, S., Borsboom, D., and Fried, E.~I. (2016).
\newblock Estimating psychological networks and their stability: A tutorial
  paper.
\newblock Retrieved from https://arxiv.org/abs/1604.08462.

\bibitem[Epskamp et~al., 2012]{epskamp2012qgraph}
Epskamp, S., Cramer, A. O.~J., Waldorp, L.~J., Schmittmann, V.~D., Borsboom,
  D., et~al. (2012).
\newblock qgraph: Network visualizations of relationships in psychometric data.
\newblock {\em Journal of Statistical Software}, 48(4):1--18.

\bibitem[Epskamp and Fried, 2017]{tutorial}
Epskamp, S. and Fried, E. (2017).
\newblock A tutorial on regularized partial correlation networks.
\newblock {\em arXiv preprint}, page arXiv:1607.01367.

\bibitem[Epskamp et~al., 2017]{dynamics}
Epskamp, S., Waldorp, L.~J., M\~{o}ttus, R., and Borsboom, D. (2017).
\newblock Discovering psychological dynamics in time-series data.
\newblock {\em arXiv preprint}, page arXiv:1609.04156.

\bibitem[Foygel and Drton, 2010]{foygel2010extended}
Foygel, R. and Drton, M. (2010).
\newblock Extended {Bayesian} information criteria for {Gaussian} graphical
  models.
\newblock {\em Advances in Neural Information Processing Systems},
  23:2020--2028.

\bibitem[Fried et~al., 2016]{fried2016}
Fried, E.~I., Epskamp, S., Nesse, R.~M., Tuerlinckx, F., and Borsboom, D.
  (2016).
\newblock {What are `good' depression symptoms? Comparing the centrality of DSM
  and non-DSM symptoms of depression in a network analysis}.
\newblock {\em Journal of Affective Disorders}, 189:314--320.

\bibitem[Friedman et~al., 2008]{friedman2008sparse}
Friedman, J.~H., Hastie, T., and Tibshirani, R. (2008).
\newblock Sparse inverse covariance estimation with the graphical lasso.
\newblock {\em Biostatistics}, 9(3):432--441.

\bibitem[Friedman et~al., 2014]{glasso}
Friedman, J.~H., Hastie, T., and Tibshirani, R. (2014).
\newblock {\em glasso: Graphical lasso- estimation of Gaussian graphical models
  ({R} package version 1.8)}.

\bibitem[Goldberg, 1993]{goldberg1998structure}
Goldberg, L. (1993).
\newblock {The Structure of Phenotypic Personality Traits}.
\newblock {\em American Psychologist}, 48(1):26--34.

\bibitem[Goldberg, 1990]{goldberg2001alternative}
Goldberg, L.~R. (1990).
\newblock An alternative ``description of personality'': the big-five factor
  structure.
\newblock {\em Journal of personality and social psychology}, 59(6):1216--1229.

\bibitem[Haslbeck and Waldorp, 2016]{mgm}
Haslbeck, J. M.~B. and Waldorp, L.~J. (2016).
\newblock {mgm}: Structure estimation for time-varying mixed graphical models
  in high-dimensional data.
\newblock {\em arXiv preprint}, page arXiv:1510.06871.

\bibitem[Isvoranu et~al., 2017]{isvoranu}
Isvoranu, A.~M., van Borkulo, C.~D., Boyette, L., Wigman, J. T.~W., Vinkers,
  C.~H., Borsboom, D., and {GROUP Investigators} (2017).
\newblock {A Network Approach to Psychosis: Pathways between Childhood Trauma
  and Psychotic Symptoms}.
\newblock {\em Schizophrenia Bulletin}, 43(1):187--196.

\bibitem[Kalisch et~al., 2012]{kalisch2012causal}
Kalisch, M., M{\"a}chler, M., Colombo, D., Maathuis, M.~H., and B{\"u}hlmann,
  P. (2012).
\newblock Causal inference using graphical models with the {R} package pcalg.
\newblock {\em Journal of Statistical Software}, 47(11):1--26.

\bibitem[Kossakowski et~al., 2015]{kossakowski2015}
Kossakowski, J.~J., Epskamp, S., Kieffer, J.~M., van Borkulo, C.~D., Rhemtulla,
  M., and Borsboom, D. (2015).
\newblock The application of a network approach to health-related quality of
  life ({HRQoL}): Introducing a new method for assessing hrqol in healthy
  adults and cancer patient.
\newblock {\em Quality of Life Research}, 25:781--92.

\bibitem[Kr{\"a}mer et~al., 2009]{parcor}
Kr{\"a}mer, N., Sch{\"a}fer, J., and Boulesteix, A.-L. (2009).
\newblock Regularized estimation of large-scale gene association networks using
  graphical gaussian models.
\newblock {\em BMC Bioinformatics}, 10(1):1--24.

\bibitem[Lauritzen, 1996]{lauritzen1996graphical}
Lauritzen, S.~L. (1996).
\newblock {\em Graphical models}.
\newblock Clarendon Press, Oxford, UK.

\bibitem[McCrae and Costa, 1997]{mccrae1997personality}
McCrae, R.~R. and Costa, P.~T. (1997).
\newblock Personality trait structure as a human universal.
\newblock {\em American Psychologist}, 52(5):509--516.

\bibitem[McGill et~al., 1978]{mcgill1978variations}
McGill, R., Tukey, J.~W., and Larsen, W.~A. (1978).
\newblock Variations of box plots.
\newblock {\em The American Statistician}, 32(1):12--16.

\bibitem[McNally et~al., 2015]{mcnally2015mental}
McNally, R.~J., Robinaugh, D.~J., Wu, G.~W., Wang, L., Deserno, M.~K., and
  Borsboom, D. (2015).
\newblock Mental disorders as causal systems a network approach to
  posttraumatic stress disorder.
\newblock {\em Clinical Psychological Science}, 3(6):836--849.

\bibitem[Newman, 2010]{newman2010}
Newman, M. E.~J. (2010).
\newblock {\em {Networks: an introduction}}.
\newblock Oxford University Press, Oxford, UK.

\bibitem[Olsson, 1979]{olsson1979maximum}
Olsson, U. (1979).
\newblock Maximum likelihood estimation of the polychoric correlation
  coefficient.
\newblock {\em Psychometrika}, 44(4):443--460.

\bibitem[Olsson et~al., 1982]{olsson1982polyserial}
Olsson, U., Drasgow, F., and Dorans, N.~J. (1982).
\newblock The polyserial correlation coefficient.
\newblock {\em Psychometrika}, 47(3):337--347.

\bibitem[Revelle, 2010]{psych}
Revelle, W. (2010).
\newblock {\em {psych}: Procedures for Psychological, Psychometric, and
  Personality Research ({R} package version 1.0-93)}.
\newblock Northwestern University, Evanston, Illinois.

\bibitem[Rosseel, 2012]{lavaan}
Rosseel, Y. (2012).
\newblock {lavaan}: An {R} package for structural equation modeling.
\newblock {\em Journal of Statistical Software}, 48(2):1--36.

\bibitem[Schmittmann et~al., 2013]{schmittmann2013deconstructing}
Schmittmann, V.~D., Cramer, A.~O., Waldorp, L.~J., Epskamp, S., Kievit, R.~A.,
  and Borsboom, D. (2013).
\newblock Deconstructing the construct: A network perspective on psychological
  phenomena.
\newblock {\em New Ideas in Psychology}, 31(1):43--53.

\bibitem[Tibshirani, 1996]{tibshirani1996regression}
Tibshirani, R. (1996).
\newblock Regression shrinkage and selection via the lasso.
\newblock {\em Journal of the Royal Statistical Society. Series B
  (Methodological)}, 58:267--288.

\bibitem[van Borkulo et~al., 2014]{van2014new}
van Borkulo, C.~D., Borsboom, D., Epskamp, S., Blanken, T.~F., Boschloo, L.,
  Schoevers, R.~A., and Waldorp, L.~J. (2014).
\newblock A new method for constructing networks from binary data.
\newblock {\em Scientific Reports}, 4(5918):1--10.

\bibitem[van Borkulo et~al., 2015]{van2015association}
van Borkulo, C.~D., Boschloo, L., Borsboom, D., Penninx, B. W. J.~H., Waldorp,
  L.~J., and Schoevers, R.~A. (2015).
\newblock Association of symptom network structure with the course of
  depression.
\newblock {\em JAMA Psychiatry}, 72(12):1219--1226.

\bibitem[Wasserman and Faust, 1994]{wasserman1994social}
Wasserman, S. and Faust, K. (1994).
\newblock {\em Social network analysis: Methods and applications}, volume~8.
\newblock Cambridge university press, Cambridge, UK.

\bibitem[Zhao and Yu, 2006]{zhao2006model}
Zhao, P. and Yu, B. (2006).
\newblock On model selection consistency of lasso.
\newblock {\em The Journal of Machine Learning Research}, 7:2541--2563.

\bibitem[Zhao et~al., 2015]{huge}
Zhao, T., Li, X., Liu, H., Roeder, K., Lafferty, J., and Wasserman, L. (2015).
\newblock {\em huge: High-Dimensional Undirected Graph Estimation ({R} package
  version 1.2.7)}.

\end{thebibliography}

\end{document}